\documentclass{iopart}

\usepackage{graphicx}
\usepackage{amssymb}
\usepackage{color}
\newcommand{\ket}[1]{\ensuremath{|{#1}\rangle}}

\begin{document}

\title{Relativistic quantum mechanics with trapped ions}

\author{L Lamata$^{1,2}$, J Casanova$^{1}$, R Gerritsma$^{3,4}$, C F Roos$^{3,4}$, J J Garc\'{\i}a-Ripoll$^{5}$ and E Solano$^{1,6}$}

\address{$^1$ Departamento de Qu\'{\i}mica F\'{\i}sica, Universidad del Pa\'{\i}s Vasco -- Euskal Herriko Unibertsitatea, Apartado 644, 48080 Bilbao, Spain}
\address{$^2$ Max-Planck-Institut f\"ur Quantenoptik,
Hans-Kopfermann-Strasse 1, 85748 Garching, Germany}
\address{$^3$ Institut f\"{u}r Quantenoptik und Quanteninformation, \"{O}sterreichische Akademie der Wissenschaften, Otto-Hittmair-Platz 1, A-6020 Innsbruck, Austria} 
\address{$^4$  Institut f\"{u}r Experimentalphysik, Universit\"{a}t Innsbruck, Technikerstrasse 25, A-6020 Innsbruck, Austria}
\address{$^5$ Instituto de F\'{\i}sica Fundamental, CSIC, Serrano 113-bis,
28006 Madrid, Spain}
\address{$^6$ IKERBASQUE, Basque Foundation for Science, Alameda Urquijo 36, 48011 Bilbao,
Spain}
 
 \ead{lucas.lamata@gmail.com}
 
\begin{abstract}
We consider the quantum simulation of relativistic quantum mechanics, as described by the Dirac equation and classical potentials, in trapped-ion systems. We concentrate on three problems of growing complexity. First, we study the bidimensional relativistic scattering of single Dirac particles by a linear potential. Furthermore, we explore the case of a Dirac particle in a magnetic field and its topological properties. Finally, we analyze the problem of two Dirac particles that are coupled by a controllable and confining potential. The latter interaction may be useful to study important phenomena as the confinement and asymptotic freedom of quarks. 
\end{abstract}


\maketitle

\section{Introduction}
The field of quantum simulators is one of the most rapidly growing
in quantum information science. It was Feynman~\cite{Feynman82} who stated thirty years ago that a controllable quantum device could emulate the dynamics of another quantum system exponentially faster than a classical computer. Since then, this hypothesis has been confirmed~\cite{Lloyd96} and, subsequently, intensive theoretical and experimental research has followed~\cite{Buluta09,Wunderlich09}.

Quantum simulators of quantum relativistic systems are among the most elegant and, among a few others, the most promising candidates for going beyond classical computational capabilities. In the last years, the simulation of black holes in Bose-Einstein condensates (BEC)~\cite{Garay00} and properties of the expanding universe~\cite{Alsing05,Schutzhold07} have been proposed, and an experiment on BEC sonic black hole has been realized~\cite{Lahav10}. A
relevant milestone was given by the proposal for the
simulation of Dirac equation and associated quantum relativistic
effects in a single trapped ion~\cite{Lamata07}, that was
subsequently experimentally realized~\cite{Gerritsma1}. This was the first experimental observation of the physics associated with the {\it Zitterbewegung} phenomenon, the fast quivering motion stemming from the Dirac equation and predicted by Schr\"{o}dinger  in the early days of quantum
mechanics~\cite{Schrodinger30}. Further developments produced the theoretical proposal~\cite{Casanova1} and
experimental realization~\cite{Gerritsma2} of the simulation of
Klein paradox~\cite{Klein29}, and a proposal
for simulating the Majorana equation and unphysical operations like
charge conjugation or time reversal with trapped
ions~\cite{Casanova2}.

Many other proposals and some experiments, in a wide variety of systems, have recently appeared,
as the simulation of {\it Zitterbewegung} in semiconductor quantum
wells~\cite{Schliemann05} or in graphene~\cite{Cserti06,Rusin09},
Klein paradox in graphene~\cite{Katsnelson06}, Dirac oscillator in
a trapped ion~\cite{Bermudez07a}, {\it Zitterbewegung} and Dirac physics with ultracold atoms~\cite{Vaishnav08,Merkl08,Goldman1,Zhang10,Lepori10,Witthaut11}, Klein paradox with atomic ensembles~\cite{Otterbach09}, optical {\it Zitterbewegung} in metamaterials~\cite{Wang09,Ling10},
delocalization of relativistic Dirac particles in cold
atoms~\cite{Zhu09}, photon wave function and {\it
Zitterbewegung}~\cite{Wang09b}, similarity of electron's {\it
Zitterbewegung} to the Adler-Bell-Jackiw anomaly in QED and its manifestation in graphene~\cite{Arunagiri09}, photonic analog of {\it Zitterbewegung} in binary waveguide arrays~\cite{Longhi10a}, {\it Zitterbewegung} theory in multiband Hamiltonians~\cite{David10}, classical {\it Zitterbewegung} in reduced plasma dynamics~\cite{Brizard10}, {\it Zitterbewegung} analogs in nonlinear frequency conversion~\cite{Longhi10}, experimental realization of an optical
 analog for relativistic quantum mechanics in an optical
 superlattice~\cite{Dreisow10}, relation between parity and {\it
 Zitterbewegung} and proposed simulation in trapped ions~\cite{Wang10}, {\it
 Zitterbewegung} in a magnetic field and proposal for trapped ion
 simulation~\cite{Rusin10}, Wilson fermions and axion electrodynamics in optical lattices~\cite{Bermudez10}, the Schwinger effect for a possible implementation with atoms in optical lattices~\cite{Szpak11},
  Dirac equation for cold atoms in artificial curved spacetimes~\cite{Boada11},  a theoretical analysis of cold atom simulation of interacting relativistic quantum field theories~\cite{Cirac10}, or an analysis of the photonic simulation of the quark model~\cite{Semiao11}.
  For a review of {\it Zitterbewegung} of electrons in
 semiconductors, see Ref.~\cite{Zawadzki11}.
 
 In this article, we make a step forward and analyze novel features of relativistic quantum mechanics simulations with trapped ions. We first analyze Klein paradox in 2+1 dimensions for different kinds of potentials. We address the problem of 2+1 Klein tunneling which contains novel and interesting features, like entanglement between the transmitted/reflected wave packets in the Klein dynamics and the transverse momentum wave function. We also point to the fact that more general nonlinear $V(x,y)$ potentials will already demand computational classical resources that are approaching what is currently feasible. Thus, a general 2+1 Klein dynamics will already be an interesting problem to be addressed by a quantum simulator. Then, we analyze a Dirac particle in a magnetic field, making here an emphasis on its topological properties, and show how it can be simulated with trapped ions. Finally, we study the quantum simulation of two Dirac particles coupled by a confining potential implemented in a system of three trapped ions. This model represents a simplified version of the MIT bag model of nuclear physics~\cite{Chodos74a,Chodos74b}. This semianalytic approach is still frequently used for analyzing quantum chromodynamics (QCD) in the non-perturbative regime. In this manner, we show that one could implement interesting QCD features like asymptotic freedom and confinement in a table-top experiment, controlled just by turning on and off a laser.

The paper is organized as follows. In Section \ref{Zitter31Section}, we briefly review the quantum simulation of the Dirac equation in trapped ions. In Section \ref{KleinSection}, we analyze the Klein paradox in 2+1 dimensions. In Section \ref{topologyDirac}, we study the simulation of a Dirac particle in a magnetic potential and its topological properties. In Section \ref{BagModelSection}, we address the problem of two Dirac particles coupled via a confining potential simulated with three trapped ions.  Finally, we present our concluding remarks in Section \ref{sectionConclusions}.

\section{Dirac equation simulation in trapped ions \label{Zitter31Section}}

The Dirac equation is arguably the most fundamental wave
equation~\cite{Thaller92}. It is historically considered as a significant step forward towards unifying quantum mechanics and special relativity, accurately describing the hydrogen atom spectrum, while predicting ab initio spin and antimatter. The Dirac equation acquires full significance in the context of quantum field theory and second quantization, where the number of quanta is not fixed. On the other hand, at the level of relativistic quantum mechanics, the Dirac single-particle solutions predict already intriguing phenomena. Among them, the best known are the {\it Zitterbewegung}~\cite{Schrodinger30} and Klein paradox~\cite{Klein29}.

The 3+1 Dirac equation reads
\begin{equation}
i\hbar\frac{\partial\psi}{\partial t}= {\cal H}^{3+1}_{\rm D} \psi =
(c\vec{\alpha}\cdot\vec{p}+ \beta mc^2)\psi,  \label{DiracEquation}
\end{equation}
where $\vec{\alpha}$ and $\beta$ are the Dirac matrices that obey the Clifford
algebra structure, $\{\alpha_i,\alpha_j\}=2\delta_{ij}$,
$\{\alpha_i,\beta\}=0$, $\beta^2=I_{4}$, that appear when
linearizing the expression for the relativistic energy,
$E=\sqrt{p^2c^2+m^2c^4}$.

The Hamiltonian operator in Eq. (\ref{DiracEquation}) expressed in its ``supersymmetric'' representation~\cite{Thaller92},
takes the form
\begin{eqnarray}
{\cal H}^{3+1}_{\rm D}  = \left( \begin{array}{cc}  0 & c ( \vec{\sigma} \cdot \vec{p} ) - i m c^2 \\
 c ( \vec{\sigma} \cdot \vec{p} ) + i m c^2 & 0
\end{array} \right) , \label{DiracHamiltonian}
\end{eqnarray}
where $\vec{\alpha}:= ( \alpha_x , \alpha_y , \alpha_z ) =
\mbox{off-diag}(\vec{\sigma},\vec{\sigma})$ is the velocity
operator, $\vec{\sigma}:= ( \sigma_x , \sigma_y , \sigma_z )$ are
the Pauli matrices, $\beta:=\mbox{off-diag}(-iI_{2},iI_{2})$, $c $
is the speed of light and $mc^2$ the electron rest energy.

A quantum simulation of the Dirac equation is a useful playground to
analyze different regimes and to verify in a table-top experiment
predicted phenomena like {\it
Zitterbewegung}~\cite{Lamata07,Gerritsma1} or Klein
paradox~\cite{Casanova1,Gerritsma2}. To perform just the Dirac
equation and {\it Zitterbewegung}, one considers a single trapped
ion of mass $M$ inside a Paul trap, with motional-mode frequencies
$\nu_x$, $\nu_y$, and $\nu_z$~\cite{Lamata07}. In the 3+1 case, the four-component Dirac bispinor will be codified in four internal
levels of the trapped ion, $| a \rangle, | b \rangle, | c \rangle$,
and $| d \rangle$.

For any spacetime dimensions, one should realize that the Dirac equation contains basically couplings $\sigma_ip_i$ (spin-orbit), and $mc^2\sigma_j$ (mass term). In general, one may implement the first kind of couplings, $\sigma_ip_i$, by a combination of Jaynes-Cummings (JC, also named red-sideband) and anti-Jaynes-Cummings (AJC, also named blue-sideband) interactions. The JC interaction excites one quantum of vibration while deexciting the internal state of the ion. The AJC interaction, in turn, excites one quantum of vibration while exciting the internal state of the ion. These interactions may be obtained by suitably chosen lasers, either in Raman or quadrupole-transition configurations, depending on the chosen ion.

 The resonant JC Hamiltonian can be written as
\begin{equation}
H_{\rm r} = \hbar \eta \tilde\Omega (\sigma^{+} a e^{i \phi_{\rm r}}
+ \sigma^{-} a^{\dagger} e^{-i \phi_{\rm r}}),
\end{equation}
where $\tilde \Omega$ is the coupling strength, $\sigma^{+}$
 and $\sigma^{-}$ are the raising
and lowering spin-$1/2$ operators, and $a$ and $a^{\dagger}$ are the
annihilation and creation operators associated with the
corresponding motional degree of freedom, either $x$, $y$, or $z$.
$\eta = k \sqrt{\hbar / 2 M \nu}$ is the Lamb-Dicke
parameter~\cite{Leibfried03}, where $k$ is the wave number of the
driving field. The AJC Hamiltonian reads
\begin{equation}
H_{\rm b} = \hbar \eta \tilde\Omega (\sigma^{+}
a^{\dagger} e^{i \phi_{\rm b}} + \sigma^{-} a e^{-i \phi_{\rm b}}).
\end{equation}

By properly adjusting laser phases and frequencies, one may combine
a JC and an AJC interactions to obtain the kind of couplings aimed
for, namely
\begin{equation}
H^{p_x}_{\sigma_x} = i \hbar \eta_x
\tilde\Omega_x \sigma_x (a^{\dagger}_x - a_x)  = 2 \eta_x \Delta_x
\tilde\Omega_x \sigma_x p_x,
\end{equation}
with $i (a_x^{\dagger} - a_x ) / 2 = \Delta_x \, p_x / \hbar$,
 where  $\Delta_x := \sqrt{\hbar / 2 M
\nu_x}$ is the spread in position along the $x$-axis of the ground
state harmonic-oscillator wavefunction and $p_x$ the corresponding
dimensioned momentum operator. These interactions may be addressed
to different internal levels and motional modes, in order to get the
terms in Eq. (\ref{DiracHamiltonian}) that are linear in the
momenta.

For implementing the second kind of couplings, $mc^2\sigma_j$, one may consider the carrier interaction between two internal levels of the ion. This consists of a coherent excitation of the internal
level, while leaving the motion unaffected. The carrier Hamiltonian
is
\begin{equation}
H_{\sigma} = \hbar \Omega ( \sigma^{+} e^{i \phi} + \sigma^{-}
e^{-i\phi}),
\end{equation}
that, for appropriately chosen phases, gives rise to the term $
\hbar \Omega\sigma_y$,  needed for the mass terms for 3+1
dimensions.

For the 3+1 dimension simulation~\cite{Lamata07}, notice that the Hamiltonian
\begin{eqnarray}
H_{\rm D}^{3+1} =  && 2 \eta \Delta \tilde\Omega
(\sigma^{ad}_x + \sigma^{bc}_x) p_x + 2 \eta \Delta
\tilde\Omega (\sigma^{ad}_y - \sigma^{bc}_y) p_y \nonumber \\ &&
+ 2 \eta \Delta \tilde\Omega (\sigma^{ac}_x-\sigma^{bd}_x)
p_z + \hbar \Omega (\sigma^{ac}_y+\sigma^{bd}_y) .
\label{DiracIonHamiltonian}
\end{eqnarray}
is composed of basic JC, AJC and carrier units as exposed above. When this is expressed in matrix form, in the basis $| a \rangle, | b \rangle, |
c \rangle$, and $| d \rangle$,
\begin{eqnarray}
H_{\rm D}^{3+1} \! = \! \left( \begin{array}{cc}  0 &  2
\eta \Delta \tilde\Omega ( \vec{\sigma} \cdot \vec{p} ) \! - \! i \hbar\Omega \\
2 \eta \Delta \tilde\Omega ( \vec{\sigma} \cdot \vec{p} ) \! + \! i \hbar\Omega & 0
\end{array} \right) \! , \label{DiracIonHamiltonianMatrix}
\end{eqnarray}
it coincides with the Dirac equation Hamiltonian in 3+1 dimensions, Eq.
(\ref{DiracHamiltonian}), with the equivalences for the speed of
light and rest energy,
\begin{eqnarray}
c := 2 \eta \Delta \tilde\Omega \,\,\, , \,\,\,\,\, mc^2 :
= \hbar\Omega . \label{Rosetta}
\end{eqnarray}

An important property of quantum simulations of relativistic quantum mechanics with trapped ions is that the values of the speed of light and rest energy in Eqs. (\ref{Rosetta}) may be controlled at will just by changing the laser intensities, $\tilde\Omega$ and $\Omega$. Thus, one may explore the transition from massless to massive Dirac fermions, or, equivalently, the transition from ultrarelativistic to nonrelativistic physics.

One of the most astonishing predictions of the single free-particle
solutions of the Dirac equation is the fast quivering motion called
{\it Zitterbewegung}. It is unexpected because it predicts an
oscillatory motion of a freely propagating electron. Thus, Galileo's
inertia law is not fully verified for a free relativistic electron, contrary to the Schr\"odinger equation case, and the free Dirac particle is expected to quiver around in the absence of potentials. The reason for this is the non-commutativity of the components of the velocity operator, which is given by $c\vec{\alpha}$. Thus, the marriage between quantum mechanics and special relativity seems to contradict the inertia law, at least to some extent and within the subtle realm of relativistic quantum physics. The {\it Zitterbewegung} phenomenon has not been observed so far for a real relativistic electron, given that the  predicted frequency, $\sim 10^{21}$s$^{-1}$, and amplitude, $\sim 10^{-11}$cm, are difficult to access experimentally. Moreover, its correct physical prediction and its validity has been constantly questioned in the last decades, whether we remain in the domain of relativistic quantum mechanics or quantum field theory.

The expression for the Dirac electron's position operator $\vec{r} =
( x , y , z)$ in the Heisenberg picture, derived from $d \vec{r} /
dt = \lbrack \vec{r} , H_{\rm D} \rbrack / i \hbar$, for 3+1
dimensions, reads~\cite{Thaller92}
\begin{eqnarray}
\vec{r}(t) = \vec{r} (0) + \frac{c^2
 \vec{p}}{{\cal H}^{3+1}_{\rm D}} \, t +  \frac{ic\hbar}{2} \left( \vec{\alpha} - \frac{c \vec{p}}{{\cal H}^{3+1}_{\rm D}} \right){{\cal H}^{3+1}_{\rm D}}^{-1} \left(e^{2i {\cal H}^{3+1}_{\rm D} t / \hbar}
- 1\right) , \nonumber \\
\label{Zitterposition31}
\end{eqnarray}
where the first term on the r.h.s. is the initial position, the
second is just the inertia law term, and the third one is the term
associated to the {\it Zitterbewegung}.

In the trapped ion simulation, the position of the ion mimics the
position of the simulated Dirac particle. The ionic position
operator evolution under the Dirac dynamics in Eq.
(\ref{DiracIonHamiltonianMatrix}), in 3+1 dimensions, is
\begin{eqnarray}
 \vec{r}(t)  =  \vec{r} (0) + \frac{4 \eta^2
\Delta^2 \tilde\Omega^2 \vec{p}}{H_{\rm D}^{3+1}} \, t
+  \left( \vec{\alpha} - \frac{2 \eta \Delta \tilde\Omega
\vec{p}}{H_{\rm D}^{3+1}} \right) \frac{i \hbar \eta \Delta
\tilde\Omega}{H_{\rm D}^{3+1}} \left(e^{2i H_{\rm D}^{3+1} t / \hbar}
- 1\right) ,\nonumber\\
\label{ZitterpositionIon31}
\end{eqnarray}
and the {\it Zitterbewegung}  frequency can be estimated in the
simulation as
\begin{eqnarray}
\omega_{\rm ZB} \approx  2 | \bar{E}_{\rm D} | /
\hbar \equiv 2 \sqrt{ 4 \eta^2 \Delta^2  \tilde\Omega^2 p^2_0 / \hbar^2 +
\Omega^2}  \approx  2 \sqrt{ N \eta^2
\tilde\Omega^2 + \Omega^2} ,
\end{eqnarray}
where $\bar{E}_{\rm D} \equiv \langle H_{\rm D} \rangle$ is the
average energy, $p_0$ is the average momentum for a peaked
distribution, and $N \equiv \langle a^{\dagger} a \rangle$ is the
phonon number, respectively. Additionally, one can estimate the {\it Zitterbewegung} amplitude
to be
\begin{eqnarray} R_{\rm
ZB}=\frac{\hbar}{2mc}\left(\frac{mc^2}{E}\right)^2= \frac{\eta
\hbar^2 \tilde{\Omega} \Omega \Delta}{4
\eta^2\tilde{\Omega}^2\Delta^2 p_0^2 + \hbar^2 \Omega^2} ,\label{ZBAmplitude}
\end{eqnarray}
and $R_{\rm ZB} \approx \Delta$, if $\eta\tilde{\Omega} \sim\Omega$,
which can be measurable in an experiment, as was shown in the 1+1
case in Ref.~\cite{Gerritsma1}.

\section{Bidimensional relativistic scattering simulated in trapped ions\label{KleinSection}}

The Klein paradox is another counterintuitive prediction of the relativistic quantum mechanical solutions of Dirac equation. A Dirac particle may tunnel through a steep barrier and propagate indefinitely regardless of the fact that the potential may be extended until arbitrarily long distances. The reason for this is that the positive energy electron may become, upon collision with the barrier, a negative energy electron and as such propagate freely throughout the barrier, wherever the potential barrier is $V\geq 2mc^2$. The standard explanation in quantum field theory is that there is enough energy in the system to produce a particle/antiparticle pair. 

The theoretical proposal~\cite{Casanova1} of the simulation of the Klein paradox in 1+1 dimensions was based on the Hamiltonian
\begin{eqnarray}
{\cal H}^{Kl,1+1}_{\rm D}  =  c \sigma_xp_x +m c^2 \sigma_z+\alpha x,
\label{DiracKleinHamiltonian11}
\end{eqnarray}
where $\alpha$ is the potential gradient constant. The experimental simulation of the Klein paradox relied on the manipulation of two trapped ions~\cite{Gerritsma2}. The center of mass mode together with the internal degrees of freedom of one of them simulates the positive and negative energy spinors, while the second one implements the external potential that produces the Klein paradox~\cite{Casanova1,Gerritsma2}. In order to measure the motional state of the ions in the previous simulations, the standard approach is to couple the motion to the internal states of one ion and, then, apply fluorescence detection to measure the internal state with novel techniques~\cite{Solano06,Bastin06,Santos07,Lamata07,Gerritsma1}.

Here, we propose an extension of the Klein paradox simulation to 2+1 dimensions. We will show that, for some potentials, this problem may be addressed with the techniques already developed for 1+1 dimensions. On the other hand, for arbitrary potentials in 2+1 dimensions, one is already approaching the limit of classical computational power and, in this case, a quantum simulator could already predict physics beyond current capabilities. At the end of this section, we give a heuristic analysis of this issue.

We begin by analyzing the Dirac equation in 2+1 dimensions with a linear potential along the $x$ coordinate and constant along the $y$ coordinate. The expression reads
\begin{eqnarray}
{\cal H}^{Kl,2+1}_{\rm D}  =  c \sigma_xp_x +c\sigma_yp_y +m c^2 \sigma_z+\alpha x,
\label{DiracKleinHamiltonian21}
\end{eqnarray}
where $\alpha$ is the potential gradient constant.
In order to analyze this dynamics, one can point out that $p_y$ is a constant of motion: $[{\cal H}^{Kl,2+1}_{\rm D},p_y]=0$, such that one may solve the problem for each value of $p_y$. One can realize that $c\sigma_yp_y +m c^2 \sigma_z$, with $p_y$ fixed, can be expressed as a new Pauli matrix with an effective mass coefficient,
\begin{equation}
c\sigma_yp_y +m c^2 \sigma_z=\tilde{m}c^2\tilde{\sigma}_{\tilde{y}},
\end{equation}
with $\tilde{m}c^2=\sqrt{p_y^2c^2+m^2c^4}$. Here $\tilde{\sigma}_{\tilde{y}}=n_y\sigma_y+n_z\sigma_z$ is an effective Pauli matrix in the $y-z$ plane (with a basis change in that plane), with $(n_y,n_z)=(p_y/\tilde{m}c,m/\tilde{m})$. Accordingly, the problem, for each $p_y$, is reduced to a Klein Hamiltonian in 1+1 dimensions,
\begin{eqnarray}
{\cal H}^{Kl,1+1}_{\rm D, eff}(p_y)  =  c \sigma_xp_x +\tilde{m} c^2 \tilde{\sigma}_{\tilde{y}}+\alpha x.
\label{DiracKleinHamiltonian11effective}
\end{eqnarray}

\begin{figure}[h!]
\begin{center}
\includegraphics[width=1\linewidth]{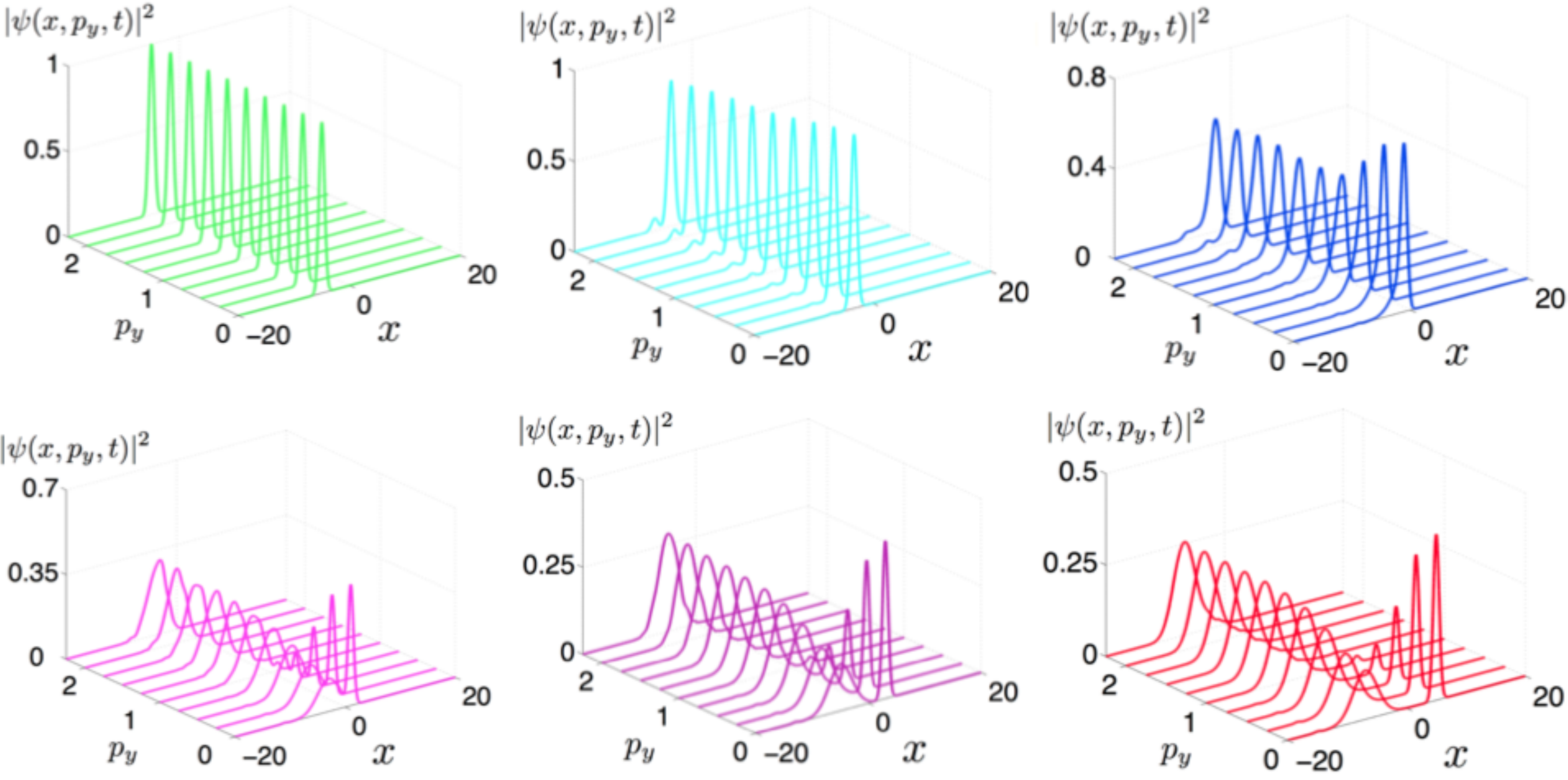}\hfill 
\end{center}\caption{Probability $| \psi(x,p_y,t)|^2$ as a function of $x$, and the parameter $p_y$ (constant of the motion), for $t=0,20,40,60,80,100$
($\hbar=c=\alpha=1$, $m=0.5$).}
                \label{Fig2}
\end{figure}

In Fig.~\ref{Fig2}, we illustrate the dependence of the Klein tunneling on the incoming momentum orthogonally to the potential gradient, $p_y.$ To do so, we simulate a wavepacket which initially has a Gaussian profile on the $x$ direction and a well defined momentum $p_y$. We plot the resulting wavefunction (normalized just in $x$ and constant in $p_y$ for the sake of comparison between different $p_y$ components), $|\psi(x,p_y,t)|^2,$ for different time snapshots $t=0,20,40,60,80,100$, in units of $\hbar=c=\alpha=1$, $m=0.5$. These plots evidence the rapid suppression of Klein tunneling for increasing $p_y.$ Given that the transmitted amplitude decreases exponentially with the effective mass $\tilde{m}$ as $\exp(- \pi \tilde{m}^2c^4/ \hbar c\alpha)$~\cite{Casanova1}, we find that already at $p_y=1$ almost all the wavefunction gets reflected.

The plots in Fig.~\ref{Fig2} admit another interpretation, based on the entanglement between the transmitted and reflected wave packets in the $x$ coordinate and the transverse momentum along the $y$ direction. In the language of quantum field theory, the electron-positron pair production along $x$ direction will be conditional to the specific value of each momentum wave packet component along $y$ direction. Thus, one could control pair production just by changing incidence angle, $p_y/p_x,$ and even perform quantum logic on pair creation conditional to the value of the transverse momentum.

Based on the behaviour at fixed values of $p_y$ we can reconstruct the dynamics of arbitrary wavepackets, fully in position space. More precisely, we can write
\begin{eqnarray}
|\psi(t)\rangle=\sum_s \int dp_y dx  e^{-it{\cal H}^{Kl,1+1}_{\rm D, eff}(p_y)/\hbar }\psi_0(x,p_y,s)|x\rangle\otimes|p_y\rangle\otimes|s\rangle,\label{DynamicsKleinPy}
\end{eqnarray}
where $\psi_0(x,p_y,s)$ is the initial state of the wavepacket, expressed in position space for the X coordinates, $\ket{x},$ momentum for the Y coordinate, $\ket{p_y},$ and spinor state, $\ket{s=0,1}.$ This reconstruction formula evidences the conditional evolution mentioned before, and in particular the entanglement between the $x$ coordinate transmitted and reflected wave packets, and the $y$ component wave packet.

\begin{figure}[h!]
\begin{center}
\includegraphics[width=1\linewidth]{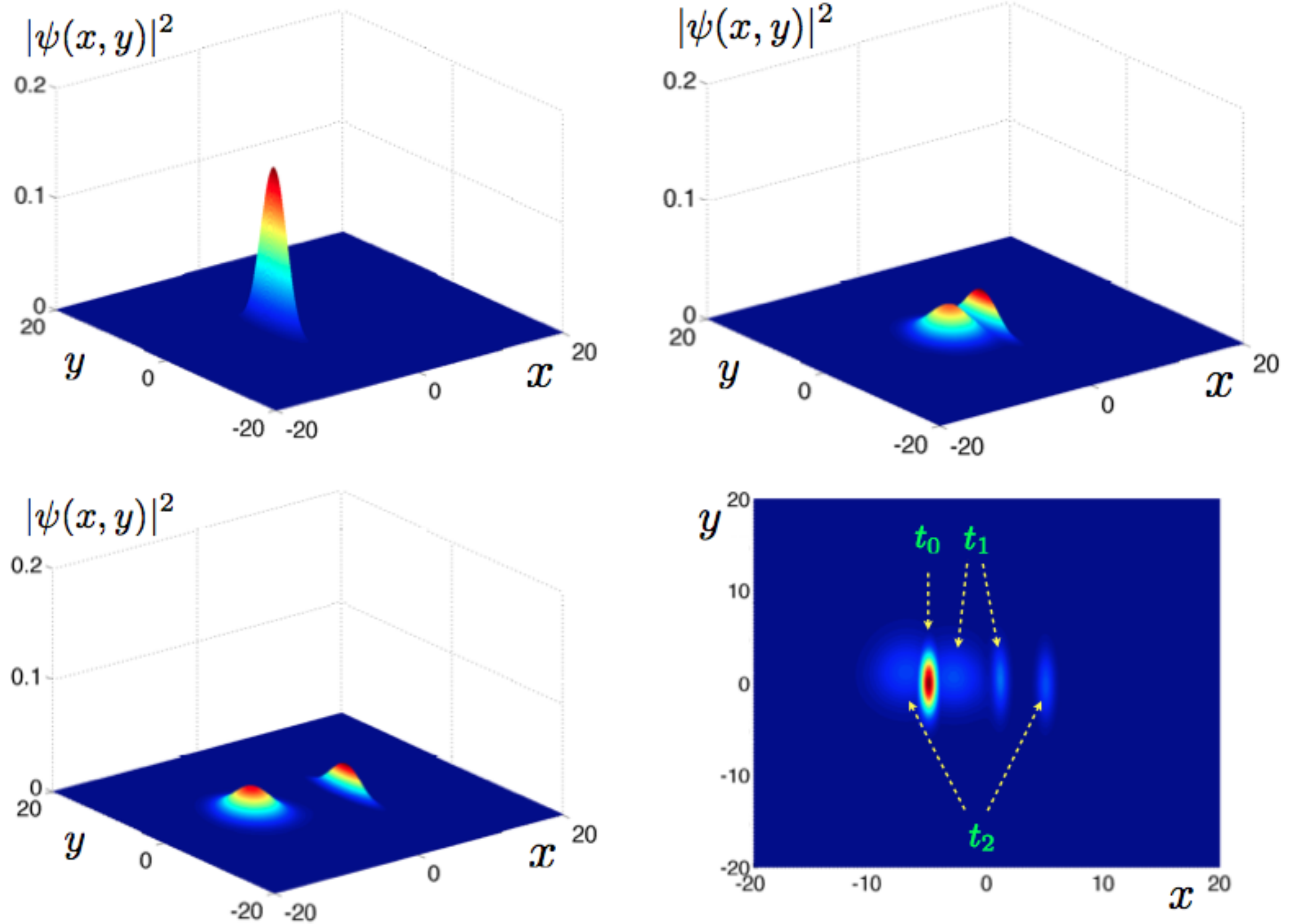}\hfill 
\end{center}\caption{Probability $| \psi(x,y)|^2$ as a function of $x$, $y$,  for $t=0, 60,100$, and a joint figure with the three times in a contour plot.
($\hbar=c=\alpha=1$, $m=0.5$).}
                \label{Fig3}
\end{figure}

In Figs.~\ref{Fig3}(a-c), we illustrate solutions obtained using the reconstruction procedure (\ref{DynamicsKleinPy}), plotting the wavefunction on position space, $|\psi(x,y,t)|^2,$ at three instants of time $t=0,60,100.$ In addition, to ease comparison, Fig.~\ref{Fig3}d shows a plot that combines all three density distributions. The most obvious effect is the squeezing of the transmitted and the broadening of the reflected wave functions after the collision with the potential barrier. To understand this phenomenon we must realize that the energy band curvature induced by $\tilde{m}(p_y)$ is associated to a momentum-dependent group velocity
\begin{equation}
  v_g(p_x) \sim \frac{d}{dp_x} E = \frac{c^2 p_x}{\sqrt{c^2p_x^2 + \tilde{m}(p_y)^2c^4}},
\end{equation}
which in turn leads to a spreading of wavepackets in position space. This spreading is most relevant for small values of $p_x,$ that is when the particle crashes against the potential barrier. At this point, which is when the particle splits into a transmitted and a reflected component, a narrow band of $p_x$ components will be filtered and become part of the antiparticle branch, moving on with very little spreading and almost uniform group velocity. The reflected component, on the other hand, will get to see the whole curvature of the energy band and become very broad very quickly, until it reaches relativistic velocities $(v_g \simeq c)$ and stops spreading.
  
For performing the protocol, we envision to follow the lines developed for the 1+1 case~\cite{Casanova1,Gerritsma2} and use two trapped ions. One of them, together with two motional modes, will codify the spinor, including the $x$ and $y$ dependence of the wave packet. The other one will produce the linear potential in $x$. To be more specific, we consider two trapped ions trapped in a linear Paul trap, with two motional modes, with frequencies $\nu_{\rm cm}=\nu$ and $\nu_{\rm st}=\sqrt{3}\nu$. To implement the Dirac-Klein dynamics given by Eq. (\ref{DiracKleinHamiltonian21}), we proceed in an analogous way as in Refs.~\cite{Lamata07,Gerritsma1,Casanova1,Gerritsma2} and use red and blue sidebands for each of the terms $c\sigma_ip_i$, $i=x,y$, as well as an AC-Stark shift for term $mc^2\sigma_z$. The $\alpha x$ potential term will be implemented with a red and blue sidebands, with appropriate phases, applied to the second ion~\cite{Casanova1,Gerritsma2}.  The corresponding trapped-ion Hamiltonian will read
 \begin{eqnarray}
H^{Kl,2+1}_{\rm D} =  2 \eta_{\rm cm} \Delta_{\rm cm} \tilde{\Omega}_{\rm cm}
\sigma_{x,1} p_{\rm cm} +  2 \eta_{\rm st} \Delta_{\rm st} \tilde{\Omega}_{\rm st}
\sigma_{y,1} p_{\rm st} + \hbar \Omega \sigma_{z,1} \nonumber\\+ \frac{ \eta_{\rm cm}\tilde{\Omega}_0}{ \Delta_{\rm cm}}\sigma_{x,2} x_{\rm cm},
\label{DiracIonKleinConstantyHamiltonian21}
\end{eqnarray}
which coincides with Eq. (\ref{DiracKleinHamiltonian21}) when one performs the analogies $2 \eta_{\rm cm} \Delta_{\rm cm}  \tilde{\Omega}_{\rm cm}=2 \eta_{\rm st} \Delta_{\rm st} \tilde\Omega_{\rm st}\leftrightarrow c$, $\hbar \Omega\leftrightarrow mc^2$, and $ \eta_{\rm cm} \tilde\Omega_0/ \Delta_{\rm cm} \leftrightarrow \alpha$, when ion 2 is initialized in the eigenstate $+$ of $\sigma_{x,2} $.

For measuring the wave packet after performing the quantum simulation, one would proceed by mapping the motional state to the internal degrees of freedom of ion 2, and subsequently detecting the internal state by fluorescence detection, as exposed in Refs.~\cite{Solano06,Bastin06, Santos07,Lamata07,Gerritsma1}. In these references, it was shown how to perform tomography of the ion motional state in an efficient way.

We point out that, when considering a nontrivial potential $V(x,y)$ in  $x$ and $y$ coordinates, say $V(x,y)=\alpha_x x^2+\alpha_y y^2$, which could be implemented with three ions in the dispersive limit, its simplification into a set of 1+1 problems cannot be done anymore, and the full bidimensional problem has to be considered. This means that, for reasonable sizes of motional Hilbert spaces of around 200 phonons per mode, one would have a Hilbert space dimension of $2\times 200$ for the 1+1 dimensional case. On the other hand, for the 2+1 case considered here, one would have a Hilbert space dimension of $8\times 10^4.$  The size of the states and the size of the associated Hamiltonian matrix force us to adopt particularly efficient schemes in the simulation, such as using finite differences or Fourier representations for the Hamiltonian and Trotter  methods for the unitary evolution. However, none of these techniques are without errors. They are severely affected by discretizations of time, and also by the boundary conditions, and even the simulations presented here are well within the limits of what can be accurately simulated, both in time and in space. These are the reasons why an actual experimental simulation of the 2+1 Klein scattering may have such an important value: without these constraints ---memory, computational power, narrow boundary conditions---, it allows for an independent and perhaps more reliable verification of our predictions.
  
  \section{Simulation of Dirac particles in magnetic fields}
\label{topologyDirac}

So far, we have seen examples in which the trapped-ion simulation may be used to learn more about the Dirac equation, but the converse is also true: working in the quantum simulations may offer a new perspective on quantum optical problems. In particular, in this section, we will relate the simulation of Dirac particles in magnetic fields with the Jaynes-Cummings model, and use this to shed light on the topological features of the eigenstates of both models ---the simulated one and the one used to implement the simulation.

The topological properties of free Dirac fermions has become a very active research area in the last years~\cite{Hasan10}. Roughly, in this context the energy band of a 2D free Dirac fermion can be seen as a mapping from momenta, $(k_x,k_y)$ to a particular spinor state on the sphere, $\vec{S} \in \mathcal{S},$ obtained from the Dirac wavefunction. The topologically nontrivial models are those for which this mapping winds up one or more times on the same sphere. The intriguing result is that such models, when embedded on surfaces with boundaries, present robust topologically protected ``edge states'', that is states that may transport charge or spin even when the bulk is placed in an insulating state (for instance with a large mass).

We will see that some of these ideas may be cast in the context of trapped-ion simulations. For that, let us recall the model of a Dirac particle moving on a uniform two-dimensional magnetic field~\cite{Bermudez07a,Bermudez07},
\begin{equation}
  H = c\sigma_x p_x + c \sigma_y [p_y - e A(x,y)] + mc^2 \sigma_z.
\end{equation}
We may choose an axial gauge in which $A(x,y) = B x.$ In this setup, the momentum along the $y$ direction becomes a constant of motion, $[p_y, H]=0.$ Changing units so that $eB=1,$ and displacing the ``x'' coordinate by an appropriate amount, $x + p_y\to x,$ we may rewrite the previous model as a simple Jaynes-Cummings model with a detuning,
\begin{equation}
  \label{eq:2}
  H = c \left ( \begin{array}{cc}
      mc & p_x + i x \\ p_x -i x & - mc \end{array} \right )
  = c \sqrt{2} (\sigma^+ a + a^\dagger \sigma^-) + mc^2 \sigma_z.
  \label{eq:mag-field}
\end{equation}
This model has a ladder of discrete energy eigenstates,
\begin{equation}
 E_{\pm n} = \pm \sqrt{ 2 c^2 n^2 + m^2 c^4 },
\end{equation}
the so called Landau levels. The wavefunctions of these levels are confined along ``x'' and form plane waves along the ``y'' direction
\begin{equation}
 \Psi_{\pm n}(x,y) = \frac{1}{\sqrt{2}}\left(\begin{array}{c}
   \alpha_{\pm n}\phi_n(x) \\ \beta_{\pm n}\phi_{n-1}(x)\end{array}\right) e^{ip_y y},
\quad n = 1,2,\ldots\label{eq:landau}
\end{equation}
where $\phi_n(x)$ is the $n$-th level of a quantum harmonic oscillator and the coefficients $(\alpha_{\pm n},\beta_{\pm n})$ are the corresponding eigenvectors of the matrix
\begin{equation}
  H_n = c \left ( \begin{array}{cc}
      mc & i \sqrt{2n} \\ -i\sqrt{2n}& - mc \end{array} \right ).
\end{equation}

\begin{figure}[t!]
  \centering
  \includegraphics[width=\linewidth]{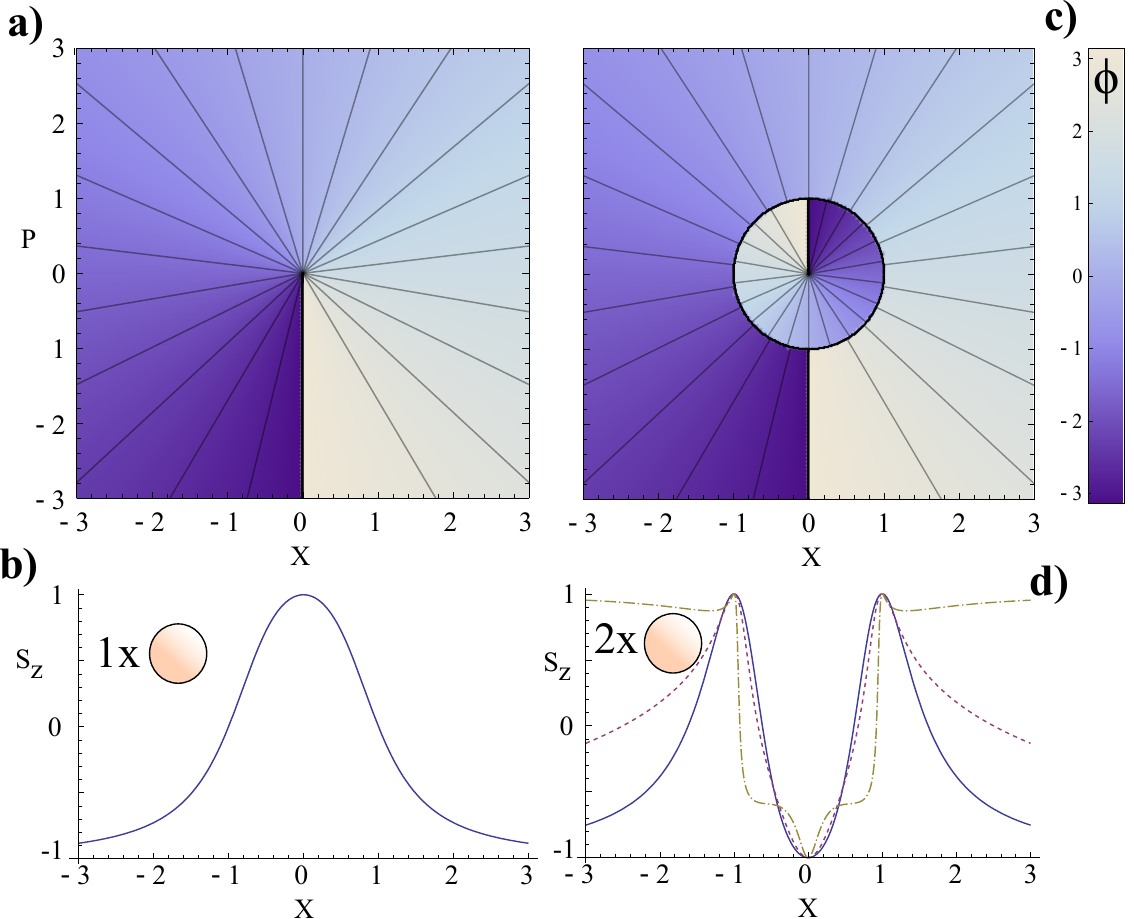}
  \caption{Wigner function of the $n=0$ and $n=1$ Landau levels, (a-b) and (c-d) respectively. The upper row shows the angle on the $\{S_x, S_y\}$ plane, $\tan^{-1}(\langle S_y\rangle/\langle S_x\rangle),$ while the lower row plots the $\langle S_z\rangle,$ along a $P=0$ cut. Note how the $n=0$ case implements a simple covering of the sphere, going from the north to the south poles, at $x=0$ and $x=\infty,$ while the $n=1$ level covers the sphere twice, reaching the South pole at $x=0$ and $x=\infty.$}
  \label{fig:wigner}
\end{figure}

At this point, instead of focusing as usual on the topology of energy bands, we will try to understand the topological properties of these individual states. For that we introduce the Wigner function of the spinor, given by
\begin{equation}
  W(\vec{x},\vec{p}) = \frac{1}{\pi\hbar}
  \int \Psi(\vec{x}+\vec{s}/2) \Psi^\dagger(\vec{x}+\vec{s}/2)
  \exp(-i \vec{s}\vec{p}/\hbar) d^2s.
\end{equation}
This quasiprobability distribution may be interpreted as a mapping from phase space $(\vec{x},\vec{p})$ to points on the Bloch sphere, $\vec{S}\in\mathcal{S},$
\begin{equation}
  \vec{s}(x,p_x) = \frac{1}{N} \mathrm{Tr}\{W(x,p_x)\vec{\sigma}\},
  \label{eq:vector}
\end{equation}
where $N = \Vert\mathrm{Tr}\{W\vec{\sigma}\}\Vert$ is a normalization factor and we have dropped the $y$ coordinate.

Figure~\ref{fig:wigner} shows the Wigner function of the two simplest Landau levels, $n=\pm 1$ and $n=\pm 2,$  decomposed into two quantities, the polar angle $\phi = \tan^{-1} (s_y / s_x),$ and the separation from the equator, $s_z.$ For the $n=0$ state we see that the pseudospin points North in the center of the phase space and rotates to the South at $r = x^2+p^2\to\infty.$ This is equivalent to a single covering of the sphere, or skyrmion, with a topological charge\footnote{We do not take the orientation into account when computing this number, and instead define the charge as the absolute number of coverings.} $\nu = 1$. To achieve greater complexity we have to increase the number of quanta, moving on to state $n=2,$ which performs a double covering of the sphere. This is done via an extended singularity of the angle $\phi$ at radius $r=1,$ which is when the spin points North. In general we have verified that the number of coverings increases linearly with the number of quanta, $\nu = \pm n,$  leading to a family of topological defects with unprecedented richness.

It may seem that these topological defects are mere artifacts of the Dirac or Jaynes-Cummings equations and that under normal circumstances, dephasing, spontaneous decay or other perturbations, will rapidly disappear. However, this is not the case. Pure dephasing, for instance along the ``z'' direction, is equivalent to a shrinking of the transverse components
\begin{equation}
  \vec{s}(t) = (s_x e^{-\gamma t}, s_y e^{-\gamma t}, s_z).
\end{equation}
This map contracts the total surface in Bloch space, which now looks closer to a rugby ball instead of a sphere. However, after suitable rescaling via a new normalization factor in Eq.~(\ref{eq:vector}), one finds that the total solid angle covered is the same and the charge, $\nu,$ is not modified.

It is even more surprising to see that spontaneous emission has also a mild effect on the winding number. Applying the positive map that corresponds to this process onto the ion state, we see that the winding on the $\{S_x, S_y\}$ plane remains unaffected by the same reasons as before, and the only change is a shift of the weights of the different harmonic oscillator eigenstates, or equivalently $s_z.$ In Fig.~\ref{fig:wigner}d, we summarize that process for the $n=2$ Landau level, showing how dissipation tries to ``unwrap'' the sphere, removing one layer and falling down to a lower winding number. Note how, despite the effort in unwrapping the sphere, the defects remained pinned and the total number constant even for long times. One may consider a final form of losses: total dissipation in the form of friction in ``x'' or ``p''. In either case, we expect population being transferred down the ladder of harmonic oscillator states, which this time would lead to a continuous unwrapping of the sphere $\nu = n, n-1, n-2\ldots 0.$

Regarding the experimental implementation, the Hamiltonian for the Dirac particle in the field has, as we have shown, a direct reinterpretation in terms of a detuned Jaynes-Cummings model (\ref{eq:mag-field}), which makes its simulation rather straightforward. Measurement of the Wigner function is also possible using standard trapped ion techniques \cite{Leibfried03}, but now the reconstruction has to be performed without tracing out the internal state of the ion. To do this will require using an ancillary ion for measurement and a different ion to implement the spinor. The ancillary ion will remain in the ground state until the end of the experiment, when it will be used to gather statistics about the vibrational mode that implements the Jaynes-Cummings Hamiltonian~\cite{Leibfried03}. The novelty is that this statistics will have to be performed conditionally on the spinor-ion, which should be simultaneously measured in one of the $\sigma_x,\,\sigma_y$ or $\sigma_z$ basis, as it is usual for wavefunction reconstruction. The resulting statistics and accuracy should be more than enough to allow for accurate reconstruction of the $\nu=1$ and $\nu=2$ skyrmions.

\section{Two Dirac particles interacting via a classical potential: Simulation of an analogue of MIT bag model}
 \label{BagModelSection}

The quantum field theory for the strong interaction, known as quantum chromodynamics (QCD), accurately predicts the behavior of nuclei, hadrons, and quarks and gluons. On the other hand, for low energies and long distances the theory becomes non-perturbative (because the coupling becomes too large) such that Feynman-diagram expansion cannot be applied.

Many decades ago, some effective theories were developed that qualitatively and quantitatively described the physics of the strong interaction even in the non-perturbative regime. Among those theories, arguably, the best known is the MIT bag model~\cite{Chodos74a,Chodos74b}. In its simplest version, it consists of two or more Dirac particles contained inside a volume, a {\it bag}, whose energy grows linearly with the volume. In this way, this model can simulate behavior like asymptotic freedom and confinement: when the fermions go far away from each other, the energy grows steadily and there is a tendency to come back to the original position, i.e., confinement takes place. On the other hand, when the fermions are nearby, the potential energy becomes negligible and the fermions behave as if they were free, i.e., asymptotic freedom holds. Although these models can be solved in a classical computer, and are phenomenological, they have quite a good agreement with experiments and predictive power. Even nowadays, this kind of methods, either with color charge or without it, string models, etc.,  are the basic tools to analyze QCD at low energies in the nonperturbative regime, with semianalytic tools (without resorting to lattice gauge theory, which requires huge computational power)~\cite{Kurkela10}. We believe it would be interesting to implement analogues of the MIT bag model in a quantum simulator, even if it is just for reproducing the physics of quark confinement and asymptotic freedom in a table top tunable experiment. In addition, quantum simulations of these models with state-of-the-art technology could already go beyond what can be computed classically (for large motional Hilbert space dimensions). In this Section, we propose the simulation of a simplified analogue of MIT bag model in 1+1 dimensions. Thus, we propose a simulation of two Dirac equations, one for each Dirac particle, that are coupled by a potential which grows with the distance, in our case quadratically for the sake of experimental feasibility. This model could describe a quark and antiquark coupled by gluons inside a meson, and cannot be solved analytically. By appropriately tuning the lasers in an experiment, one could perform phase transitions between the asymptotically free and the confined meson phases, like the one that supposedly took place in the early Universe. 

The model we aim to simulate is
\begin{eqnarray}
H_{\rm ABM}=c\sigma_{x,1}p_1+c\sigma_{x,2}p_2+mc^2\sigma_{y,1}+mc^2\sigma_{y,2}+V_0 (x_1-x_2)^2.
\end{eqnarray}
In order to simulate two Dirac particles (e.g., quark-antiquark constituting a meson), each obeying a 1+1 Dirac
equation and with a potential that grows with the separation between the particles, we envision to use three ions: two of them will
codify the positive and negative energy spinors of the two Dirac
particles, while the third one will be used to generate the
attractive potential among them. The normal modes for a three ion
system with equal mass and charge are
\begin{eqnarray}
Q_{cm} & = & \frac{1}{\sqrt{3}}(x_1+x_2+x_3),\;\; P_{cm}=\frac{1}{\sqrt{3}}(p_1+p_2+p_3),\\
Q_r & = & -\frac{1}{\sqrt{2}}(x_1-x_3),\;\; p_r=-\frac{1}{\sqrt{2}}(p_1-p_3),\\
Q_3 & = & \frac{1}{\sqrt{6}}(x_1-2x_2+x_3),\;\; P_3=\frac{1}{\sqrt{6}}(p_1-2p_2+p_3).
\end{eqnarray}
We will consider the two first modes to codify the two free Dirac
equations plus the potential. By considering the Hamiltonian
\begin{eqnarray}
\nonumber H_{\rm ABM,sim} & = & 2 \eta_{cm}\Delta_{cm}\tilde{\Omega}_{cm}(\sigma_{x,1}-\sigma_{x,3})P_{cm}+2\eta_r\Delta_r\tilde{\Omega}_r(\sigma_{x,1}+\sigma_{x,3})p_{r}
\\&&+\hbar\Omega\sigma_{y,1}+\hbar\Omega\sigma_{y,3}+\frac{\hbar\eta_{cm}\Omega_3}{\Delta_{cm}}\sigma_{x,2}Q_{cm}+\hbar\Delta_3\sigma_{z,2},
\end{eqnarray}
we simulate an equivalent dynamics to two free Dirac equations with a potential in the relative coordinate,
\begin{eqnarray}\label{BagModelHamilt}
&&H_{ABM}=c\sigma_{x,1}p_1+c\sigma_{x,3}p_3+mc^2\sigma_{y,1}+mc^2\sigma_{y,3}+V_0 (x_1-x_3)^2\\&&=
c(\sigma_{x,1}-\sigma_{x,3})\tilde{p}_{r}+c(\sigma_{x,1}+\sigma_{x,3})\frac{\tilde{P}_{cm}}{2}
+mc^2\sigma_{y,1}+mc^2\sigma_{y,3}+V_0 \tilde{x}_r^2,\nonumber
\end{eqnarray}
where $\tilde{p}_r=(p_1-p_3)/2$, and $\tilde{P}_{cm}=p_1+p_3$ are
the relative and center of mass momenta of the system of two
particles that we want to simulate, and $\tilde{x}_r=x_1-x_3$ is the
relative coordinate. To make the analogy complete, one has to make
the substitutions $P_{cm}\leftrightarrow \tilde{p}_r$,
$p_{r}\leftrightarrow \tilde{P}_{cm}/2$, $Q_{cm}\leftrightarrow
\tilde{x}_r$, with $2\eta_{cm}\Delta_{cm}\tilde{\Omega}_{cm}=2\eta_r\Delta_r\tilde{\Omega}_r=c$, $\hbar\Omega=mc^2$, and the potential is obtained in the dispersive limit,
for $\hbar\Delta_3\gg \hbar\eta_{cm}\Omega_3\langle a_{\rm cm}^\dag + a_{\rm cm}\rangle$, considering ion $2$ in the $+$
eigenstate of $\sigma_{z,2}$, and with
$V_0=(\hbar\eta_{cm}\Omega_3/\Delta_{\rm cm})^2/\hbar\Delta_3$.
\begin{figure}[t]
\begin{center}
\includegraphics[width=\linewidth]{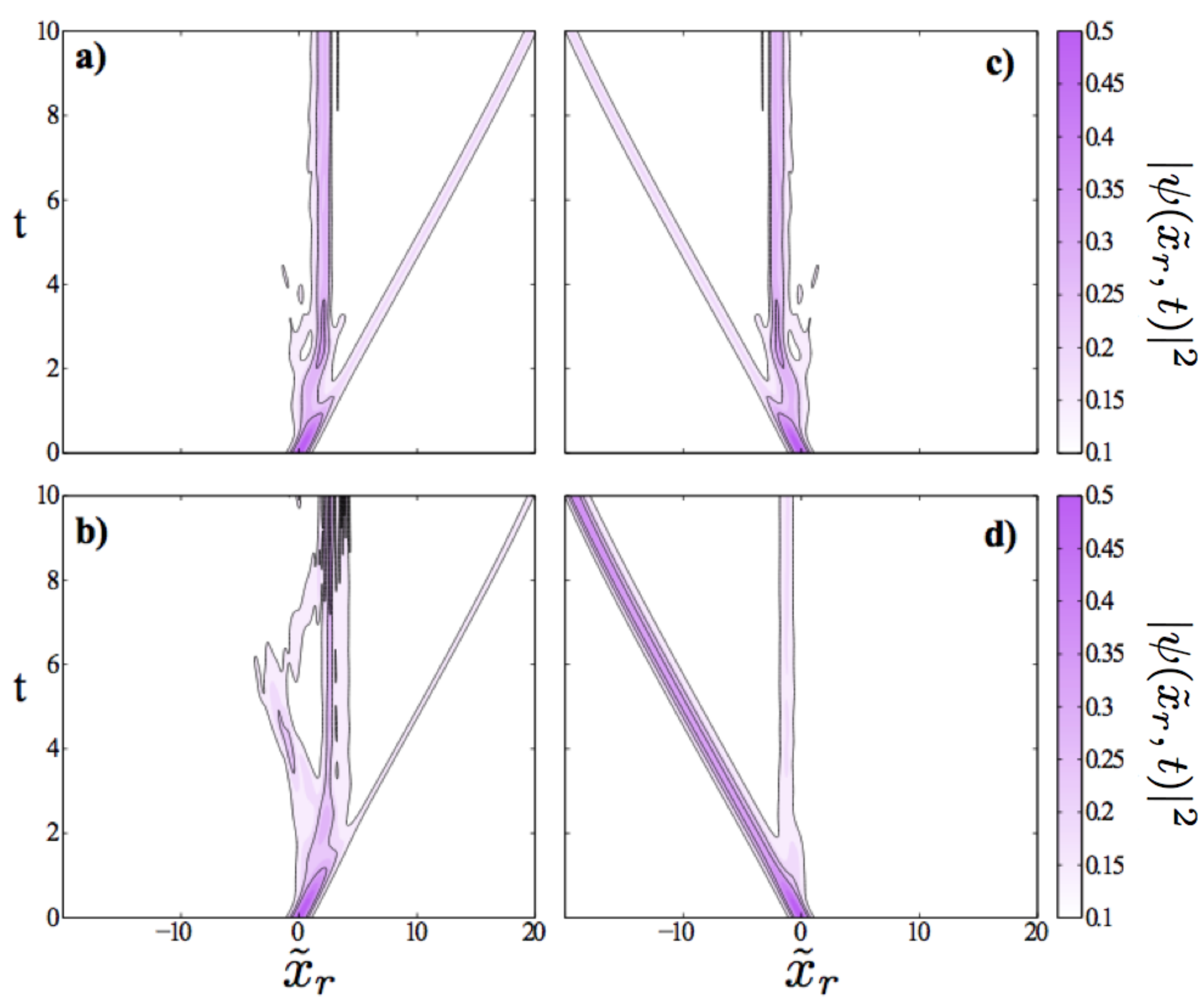}
\end{center}\caption{Probability $|\psi(\tilde{x}_r,t)|^2$ as a function of $\tilde{x}_r$ and $t$, for the two Dirac fermion system as given by Eq. (\ref{BagModelHamilt}), for $\tilde{P}_{cm}=2$ (a constant of motion), $mc^2=1$, and $V_0=0.5$, for the initial values (a) $\langle \tilde{p}_{r}\rangle=0,\,\Pi=+1,$ (b) $\langle \tilde{p}_{r}\rangle=2,\ \Pi=+1,$ (c) $\langle \tilde{p}_{r}\rangle=0,\,\,\Pi=-1,$ and (d) $\langle \tilde{p}_{r}\rangle=2,\,\Pi=-1$, with $\Pi$ the initial value of the operator $\hat{\Pi}=\frac{1}{2}(\sigma_{x,1}-\sigma_{x,3})$ on the spinor state.}
\label{FigBagModel}
\end{figure}

With this simulation, we may analyze confinement and asymptotic freedom by playing with $V_0$ potential. One would expect in principle, though predictions are complicated in systems with no analytical solutions, to have two independent dynamics. One of them will happen for the case in which the fermions are nearby ($\tilde{x}_r\simeq 0$) and the momentum $|\tilde{p}_{r}|$ is small (asymptotic freedom). And another more complicated dynamics, coupling the two fermionic motions, for those cases in which the two fermions fly away ($\tilde{x}_r$ large) (confinement). Additionally, when $V_0\langle \tilde{x}_r^2\rangle \geq 2mc^2$, one would expect Klein tunneling to take place. In this model, we interpret this Klein tunneling as a pair creation of a new quark and a new antiquark:  when the original quark and antiquark go very far away from each other, there is energy available for creating a new antiquark (that gets bound to the original quark) and a new quark (that gets bound to the original antiquark), such that the original pair can split up without violating the color neutrality standard in QCD for free composite particles. This is the natural interpretation, given that the usual explanation for Klein tunneling in quantum field theory is the creation of a particle/antiparticle pair. In QCD, the splitting of a hadron into new hadrons is well known and takes place, for example, in the jet emission in high-energy colliders.

We have realized numerical simulations of this model, and we show our results in Fig.~\ref{FigBagModel}. There, we plot the probability $| \psi(\tilde{x}_r,t)|^2$ as a function of $\tilde{x}_r$ and $t$, for the two Dirac fermion system as given by Eq. (\ref{BagModelHamilt}), for $\tilde{P}_{cm}=2$ (a constant of motion), $mc^2=1$, and $V_0=0.5$, for the initial values (a) $\langle \tilde{p}_{r}\rangle=0,\,\Pi=+1,$ (b) $\langle \tilde{p}_{r}\rangle=2,\,\Pi=+1,$ (c) $\langle \tilde{p}_{r}\rangle=0,\,\Pi=-1,$ and (d) $\langle \tilde{p}_{r}\rangle=2,\,\Pi=-1.$ In this notation, $\Pi$ denotes the expected value of the operator $\hat\Pi=\frac{1}{2}(\sigma_{x,1}-\sigma_{x,3})$ on the initial state. We consider initial Gaussian states in $\tilde{x}_r$, normalized in this coordinate, and constant $\tilde{P}_{cm}$, that will remain unchanged during the evolution, for the sake of simplicity and ease of the numerical simulation.

Despite the fact that this problem is not analytically solvable, and it is difficult to get a clear intuitive prediction of the behavior that Eq. (\ref{BagModelHamilt}) will produce, we observe interesting features. For  $\Pi=+1$, and $\langle \tilde{p}_{r}\rangle=0$ the dynamics is quasifree, besides a small Klein tunneling appearance. This may be a signature of asymptotic freedom, because the Dirac fermions do not couple much, as seen in (a). For $\Pi=+1$ and $\langle \tilde{p}_{r}\rangle=2$, the dynamics of the two-fermion system is more involved: the wave function evolves in a more complex way for larger relative momentum, splitting up, bouncing back and getting distorted. This may be related to confinement taking place, when the two fermions try to separate, as seen in (b). On the other hand, for $\Pi=-1$ (another spinor initial state), we observe that, for finite relative momentum, $\langle \tilde{p}_{r}\rangle=2$ (d), the two fermions have a larger Klein tunneling than for zero relative momentum, $\langle \tilde{p}_{r}\rangle=0$ (c). This may  also indicate that, for large enough relative momentum, the Dirac fermion pair has enough energy to create another quark-antiquark pair and escape from each other while keeping color neutrality (d). For relative momentum equal to zero, there is little Klein tunneling (c) associated to asymptotic freedom.

Summing up, we have shown than an experiment with three ions can simulate an analogue of the well-known MIT bag model~\cite{Chodos74a,Chodos74b} for two Dirac particles in 1+1 dimensions. Despite their simplicity, such experiments would allow us to study signatures of asymptotic freedom and confinement. In addition, those experiments would easily reach simulation regimes which are beyond current numerical simulation possibilities. For example, here, we were severely limited by the problem size and complexity, preventing us from studying things like adiabatic preparation of eigenstates, a full diagonalization of the spectrum, or longer time dynamics.

\section{Conclusions}
\label{sectionConclusions}

In this article, we have shown that trapped-ion physics provides a flexible platform to simulate a number of interesting cases and effects in relativistic quantum mechanics. In particular, we have studied bidimensional relativistic scattering, topological effects of Dirac particles in magnetic fields, and two Dirac particles coupled by a potential, an analogue of nuclear physics bag models.

Summarizing, we have described the bidimensional relativistic scattering for an $x$-dependent potential, showing that this case can be solved with the help of the one-dimensional case. In addition, we showed it contains interesting novel features, like entanglement between transmitted/reflected wave packets and the transverse momentum. We have also given a heuristic analysis of the complexity of more complicated potentials that will already approach the classical computational limit of resources. We have introduced a relativistic quantum mechanics model -a Dirac particle in a magnetic field- that could shed light and establish interesting analogies with an equivalent quantum optical model, namely, the Jaynes-Cummings model with detuning. We showed a possible quantum simulation of this case, including its topological features in trapped ions. Finally, we have shown that three trapped ions together with two motional modes suffice to implement the dynamics of two Dirac particles that are coupled by a confining potential, like in the bag models of nuclear physics. This could allow the implementation in a table top experiment of quark confinement and asymptotic freedom associated to QCD, while moving from one regime to the other by switching on and off a laser. This would amount to a phenomenological simulation of the phase transition between the quark-gluon plasma and the hadronic confined phases that took place in the early universe.

In a physical implementation, strings of Ca$^+$ ions, for instance, could be used and manipulated with laser light. The spinor degrees of freedom can be encoded in long-lived electronic states of this ion, as was done before in quantum simulations of relativistic quantum mechanics \cite{Gerritsma1,Gerritsma2}. Such states can have coherence times of several ms, while coupling strengths of a few 100 kHz between the spinor states can be obtained. For each motional degree of freedom, a normal motional mode of the ion string can be used and manipulated with lasers coupling the internal and motional degrees of freedom. The implementation of the proposed quantum simulations requires the validity of the Lamb-Dicke approximation, which for Ca$^+$ ions in a trap with secular frequency of $\sim$ 1 MHz and manipulated with 729~nm laser light is well justified (Lamb-Dicke parameter $\eta \sim$ 0.05 $\ll$ 1)~ \cite{Gerritsma2} . In recent experiments \cite{Gerritsma2}, it has been shown that the motional state of the ions can be coherently manipulated while reaching states that have $>$100 phonons on average, and the coherence can be sustained for more than 10 ms. These parameters could be further improved by using, e.g., ions that are less sensitive to decoherence sources and by further decreasing the Lamb-Dicke parameter with the choice of a different beam geometry.
 
 We believe that quantum simulation is one of the most promising fields inside quantum information science. At the same time, among other research lines, quantum simulations of relativistic quantum mechanics may soon allow us to consider problems that are difficult or even impossible for classical computers. In this sense, quantum simulators will give us the opportunity to enter into unexplored regimes of the physical world. Trapped ions offer one of the most promising platforms for achieving this goal.

\subsection*{Acknowledgments}

L.L. thanks the European Commission for funding through a Marie Curie IEF grant. J. C. acknowledges support from Basque Government grant BFI08.211. J. J. G.-R. acknowledges funding from Spanish MICINN Projects FIS2009-10061 and CAM research consortium QUITEMAD S2009-ESP-1594. E. S. is grateful to Spanish MICINN FIS2009-12773-C02-01, Basque Government Grant IT472-10, SOLID and CCQED European projects.

\end{document}